\documentclass{emulateapj}

\usepackage{amsmath}

\shorttitle{Signatures of the M31-M32 Galactic Collision}
\shortauthors{M. Dierickx}

\begin{document}

\title{Signatures of the M31-M32 Galactic Collision}
\author{M. Dierickx\altaffilmark{1}, L. Blecha\altaffilmark{2,3}, and A. Loeb\altaffilmark{1}}
%\email{{\it mdierickx@cfa.harvard.edu}}
\altaffiltext{1}{Astronomy Department, Harvard University, 60 Garden Street, Cambridge, MA 02138, USA; mdierickx@cfa.harvard.edu, aloeb@cfa.harvard.edu}
\altaffiltext{2}{Einstein Fellow}
\altaffiltext{3}{Department of Astronomy, University of Maryland, CSS 1204, College Park, MD 20742; lblecha@astro.umd.edu}

%lblecha@astro.umd.edu
%301-405-6830.

\begin{abstract}
The unusual morphologies of the Andromeda spiral galaxy (M31) and its dwarf companion M32 have been characterized observationally in great detail. The two galaxies' apparent proximity suggests that Andromeda's prominent star-forming ring as well as M32's compact elliptical structure may result from a recent collision. Here we present the first self-consistent model of the M31-M32 interaction that simultaneously reproduces observed positions, velocities, and morphologies for both galaxies. Andromeda's spiral structure is resolved in unprecedented detail, showing that a rare head-on orbit is not necessary to match Andromeda's ring-like morphology. The passage of M32 through Andromeda's disk perturbs the disk velocity structure. We find tidal stripping of M32's stars to be inefficient during the interaction, suggesting that some cEs are intrinsically compact. Additionally, the orbital solution implies that M32 is currently closer to the Milky Way than models have typically assumed, a prediction that may be testable with upcoming observations.
\end{abstract}

%%%%%%%%%%%%%%%%%%%%%%%%%%%%%%
\section{Introduction}
\label{sec:intro}

Much effort has been dedicated to characterizing our ``sister galaxy'' M31 with observational surveys such as the Pan-Andromeda Archaeological Survey and the Panchromatic Hubble Andromeda Treasury. The striking star-forming ring structure of Andromeda's disk could result from orbital resonances driven by bar instability or a galactic collision.  However, because the rings are slightly offset from the disk center and differ from the expected resonance corotation radii, an off-center collision is the most likely cause \citep{braun91,gordon06, block06}.

Prominent among Andromeda's companions is the compact elliptical (cE) galaxy M32. Superimposed on Andromeda's disk, M32 has long been suspected of distorting the underlying spiral structure via a past interaction \citep{schwarzschild54,byrd76,cepa88}. A theoretical understanding of Andromeda's assembly history is lacking however, as efforts to model past satellite mergers have been hindered by unknown phase-space coordinates. While it is well known that Andromeda's line-of-sight velocity is $\sim$100~km~s$^{-1}$ greater than that of M32 \citep{mateo98}, large uncertainties remain on the two galaxies' transverse velocities \citep[e.g.][]{vdM08, sohn12}. M32 appears only $\sim$5~kpc offset from the center of M31 in projection; however, its line-of-sight distance is not tightly constrained. Indirect evidence that M32 lies in front of M31 comes from its blue color, the absence of superimposed dust clouds \citep{ford78} as well as one candidate M31/M32 microlensing event \citep{paulin-henriksson02}. Uncertain proper motions and physical separation make assigning an orbit to M32 a challenging task. Driven by newly available high-quality infrared imaging of Andromeda's ring-like structure, recent numerical works have modeled M31-M32 interactions \citep{gordon06,block06}. Simulations where M32 is treated as a point particle fail to reproduce ring-like features, and the interaction signatures dissipate within 20~Myr due to differential rotation \citep{gordon06}. The first full $N$-body simulation of an M31-M32 encounter succeeds in producing a ring in Andromeda's disk \citep{block06}, but does not match the two galaxies' observed positions and velocities \citep{davidge12}. 

A number of alternative scenarios have been proposed to explain Andromeda's morphology. For example, \citet{hammer10} successfully model key properties of M31 with a 3:1 gas-rich merger $\sim$9~Gyr ago. With distorted outer isophotes, the spheroidal galaxy NGC~205 is another close companion of M31 that shows signs of tidal interaction. A stellar stream possibly tracing out its orbit \citep{mcconnachie04} was later refuted by theoretical constraints on infall direction \citep{howley08}. NGC~205's infall velocity suggests that it is on its first pass towards Andromeda and cannot be responsible for past disruptions to M31's disk \citep{howley08}. 

The origin of the compact morphology and high surface brightness of M32-like galaxies is still debated. Because most of the known cEs tend to be found close to massive galaxies, they are often conjectured to arise through tidal stripping driven by interactions with larger neighbors \citep{faber73,bekki01, chilingarian09}. The finding of two M32 analogues with clear evidence of tidal streams by  \citet{huxor11} argues in favor of the stripping scenario. An alternative view suggests, however, that cEs are the low-mass continuation of the elliptical galaxy family \citep{wirth84,kormendy09}. This is supported by the recent finding of a ``free-flying", isolated cE \citep{huxor13}. 
The evolution of dwarf spheroidal galaxies is thought to be dictated by tides \citep[e.g.][]{mayer07,kormendy09}, with evidence that the core radius remains unaffected even in the event of extreme mass loss over multiple passages \citep{penarrubia08}. Modeling the formation of compact M32-like dwarfs thus remains challenging, as e.g. \citet{bekki01} simulate the stripping of a spiral galaxy down to a bulge resembling a cE, while \citet{choi02} observe little evolution from tidal effects. 

Our work uses a combination of full hydrodynamic simulations and test particle modeling to simultaneously reproduce the morphologies and orbits of Andromeda and M32. We describe our numerical methods in \S \ref{sec:sims}. The results from our fiducial simulation are presented in \S \ref{sec:results}, and our conclusions are summarized in \S \ref{sec:conclusions}.

%%%%%%%%%%%%%%%%%%%%%%%%%%%%%%
\section{Simulation Methods}
\label{sec:sims}

\subsection{Preliminary {\footnotesize GADGET} simulations}
\label{subsec:prelimsim}

First we conduct a preliminary suite of hydrodynamic minor-merger simulations with the smoothed-particle hydrodynamics (SPH)~/~N-body code {\footnotesize GADGET-3} \citep{springel05,springel03}. The aim is to constrain the parameter space of the M31-M32 orbit and find a plausible model for M32's progenitor galaxy, accounting for the important effects of dissipative forces. The simulations start with an M31 virial mass of $1.6\times 10^{12}$ M$_{\odot}$ and mass ratios ranging from 1:7 to 1:30. To explore the variation in ring structure created by the passage, we simulate collisions with varying incidence angles and impact parameters. From these test simulations we find that spiral structure appearing ring-like at high disk inclination forms even for inclined orbits with impact parameters of order 10~kpc, in agreement with results for more moderate mass ratio collisions \citep[e.g.][]{fiacconi12,mapelli12}. Off-center impacts are more probable and more likely to create perturbations offset from the disk center. Crucially, in each case only the first passage is strong enough to excite ring-like waves in the main disk. Later phases of the merger are excluded, as these initial disturbances dissipate within $\sim1$~Gyr. The persistence time of the spiral structure and minimum mass ratio necessary to excite waves are consistent with the flyby collisions simulated by \citet{struck11}. We conclude that M32 is presently near apocenter and on its way to a second passage through M31. 

Given this consideration, plausible progenitor morphologies for M32 are severely constrained. The combination of an intermediate halo concentration parameter $c_{\rm M32} = 8$ and a small virial mass of $8 \times 10^{10}$~M$_\odot$ matches M32's low rotation velocity, as measured by \citet{howley13}. Aiming to reproduce M32's observed half-light radius of $\sim$0.1~kpc \citep{graham02}, we test a range of masses and physical sizes for the disk and bulge and find that one passage through M31's disk is not sufficient to strip the M32 progenitor to sub-kiloparsec scales. We conclude that the progenitor must be initially compact and bulge-dominated (see also \S  \ref{subsec:m32formation}).

\subsection{Point-particle model}
\label{subsec:pointparticle}

With these initial results at hand we further narrow down the orbital parameter space with a point particle model, aiming to match the observed radial velocities and positions of the two galaxies. Given the impact parameter, inclination range and number of passages inferred from our preliminary hydrodynamic simulations, this approach allows for rapid searching of the parameter space. Here M32 is modeled as a point particle evolving in a fixed Navarro, Frenk and White dark matter potential \citep[NFW;][]{NFW97} representing M31. To provide plausible cosmological initial conditions, we initialize M32 at approximately the virial radius of Andromeda and with varying angular momentum. Assuming an NFW halo profile, we track the mass enclosed within M32's tidal radius to evaluate tidal stripping and dynamical friction in time.

For a satisfactory orbit we require the M31-M32 distance to be within the error bars on their relative distance \citep[$\sim$120~kpc; e.g.][]{freedman89,choi02}. We also require trajectories to be mostly radial with respect to the Milky Way \cite[e.g.][]{vdM12II} by selecting total relative M31-M32 velocities between 90 and 110~km~s$^{-1}$. Because Andromeda's two main rings are offset from the galaxy center by 0.5-1~kpc \citep{block06}, we select an orbit with an intermediate impact parameter of $\sim10$~kpc to resimulate at high resolution with {\footnotesize GADGET-3}. As the orbital match between the approximate semi-analytic model and the SPH code is imperfect, we then tune the orbital solution with repeated SPH runs at intermediate resolution to match the observed M31-M32 configuration.

%%%%%%%%%%%%%%%%%%%%%%%%%%%%%%
\section{Results}
\label{sec:results}

\begin{deluxetable}{llll}
\tablewidth{0pt}
\tablecaption{Initial conditions for fiducial model.}
%\tablehead{\multicolumn{3}{c}{Andromeda (M31) parameters}}
\tablehead{Parameter & Description & Andromeda (M31) & M32}
\startdata
$M_{\rm 200}$ & Virial mass (M$_{\odot}$) & $1.6 \times 10^{12}$ & $8.0 \times 10^{10}$ \\
$R_{\rm 200}$ & Virial radius & 185 kpc & 67 kpc\\
$c$ & Halo concentration & 12 & 8\\
$m_{\rm b}$ & Bulge mass (M$_{\odot}$) & $2 \times 10^{10}$  & $8 \times 10^{8}$  \\
$m_{\rm d}$ & Disk mass (M$_{\odot}$) & $8\times 10^{10}$ & $8 \times 10^{8}$  \\
$f_{\rm g}$ & Disk gas fraction & 0.1 & 0.033\\
$a$ & Bulge scale length & 1.8 kpc & 0.25 kpc \\
$R_{\rm d}$ & Disk scale length & 5.5 kpc & 0.25 kpc \\
$c_{\rm 0}$ & Disk scale height & 1.14 kpc & 0.25 kpc \\
\hline \hline
\enddata
\tablecomments{M31 parameters are adapted from \citet{widrow05}. M32 parameters are chosen based on test simulations. We use standard $\Lambda$CDM cosmological parameters values of $H_{0} = 70$~km~s$^{-1}$~Mpc$^{-1}$ and $\Omega_{\rm m} = 0.27$.}
\label{tab:params}
\end{deluxetable}

% Snapshot mosaic figure
\begin{figure*}[hbt]
\begin{center}
\includegraphics[scale=0.58]{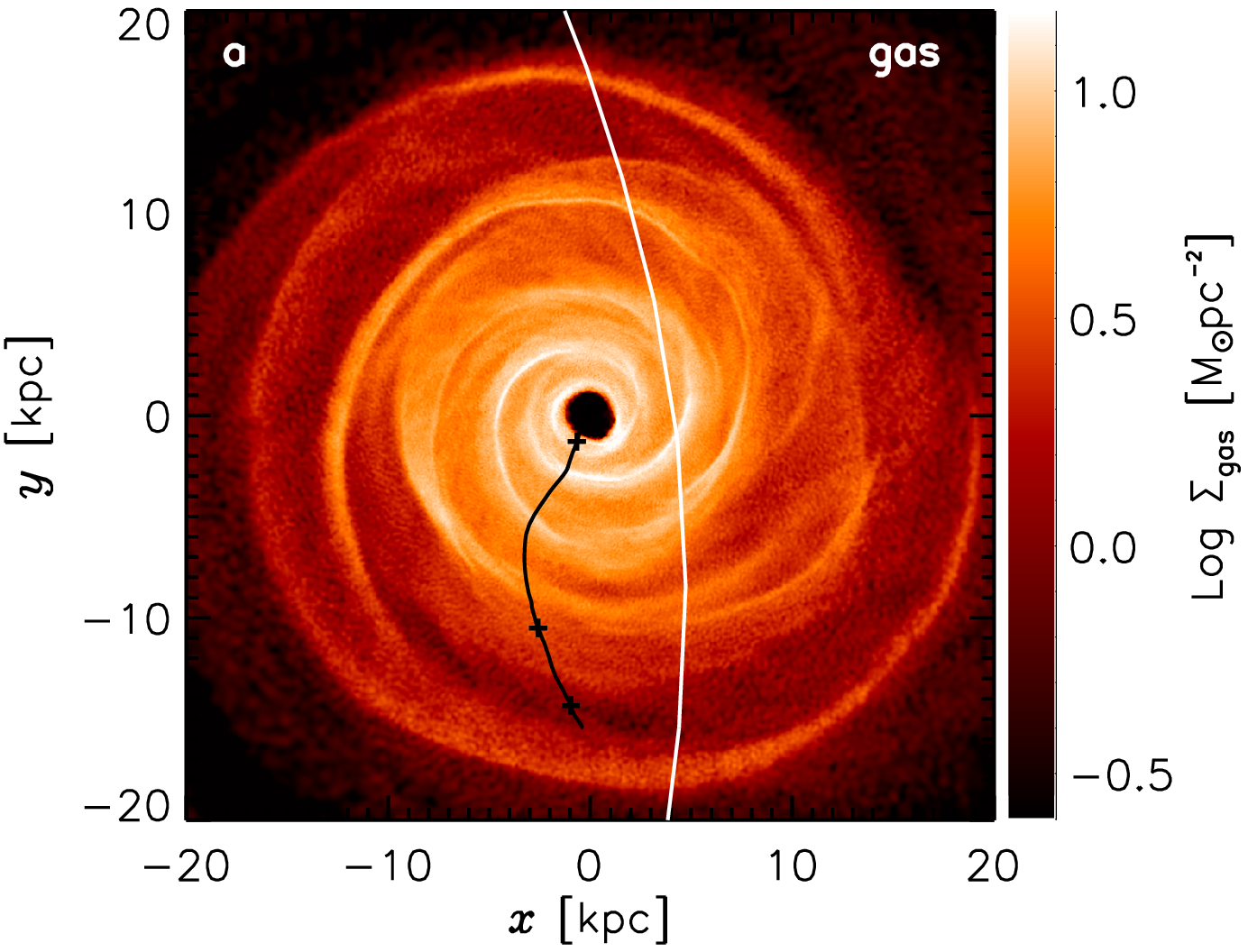} 
\includegraphics[scale=0.58]{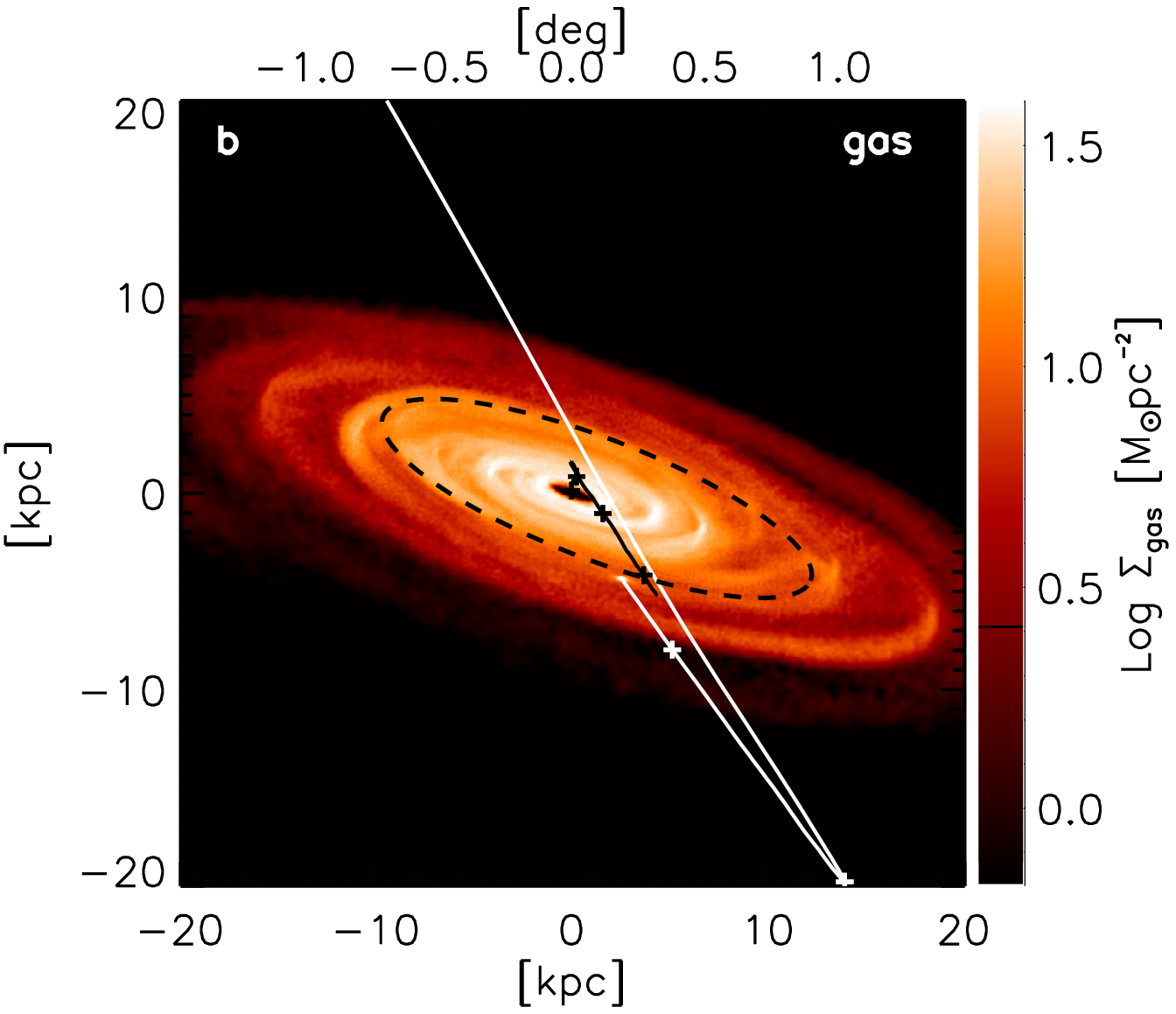}
\includegraphics[scale=0.58]{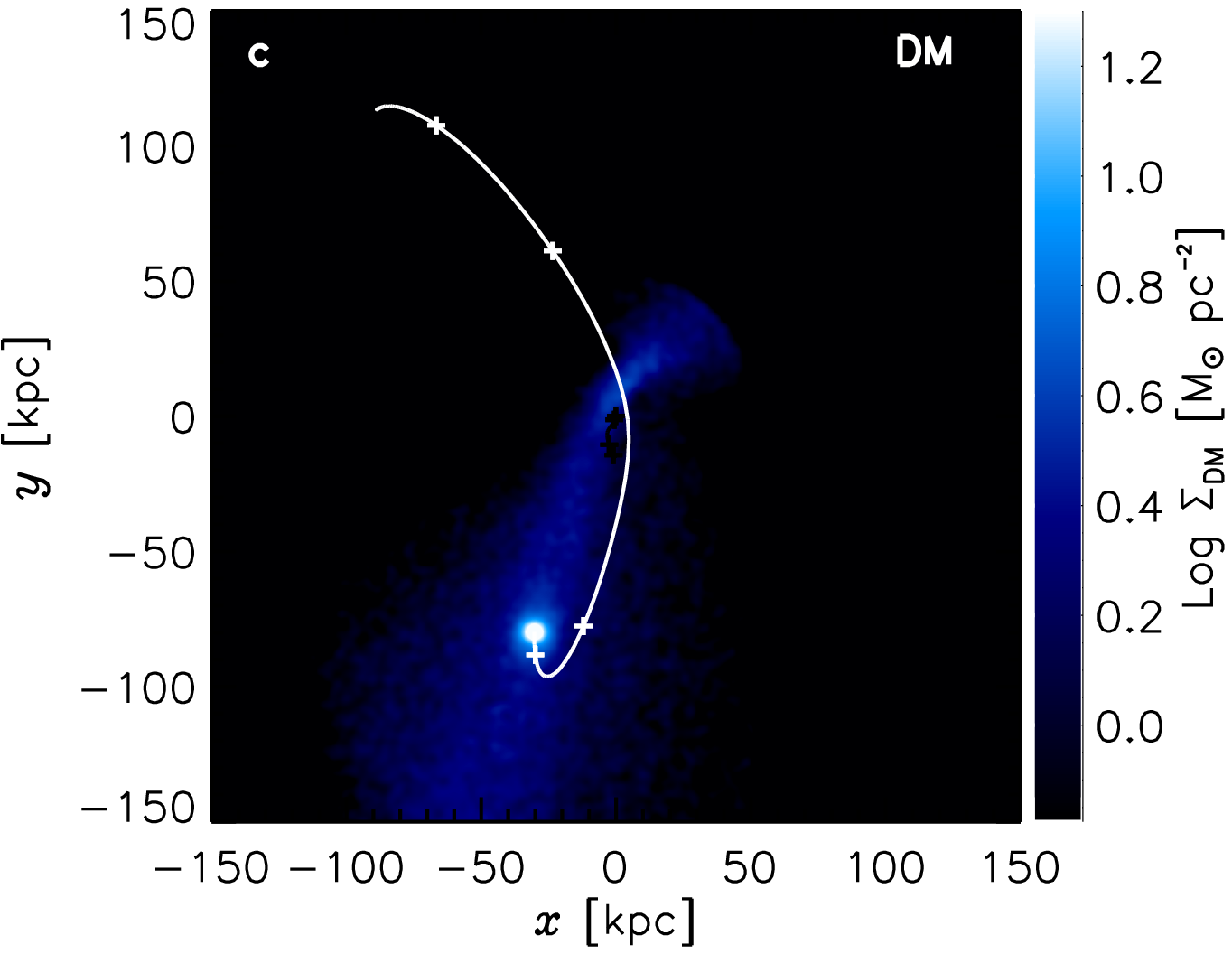} 
\includegraphics[scale=0.58]{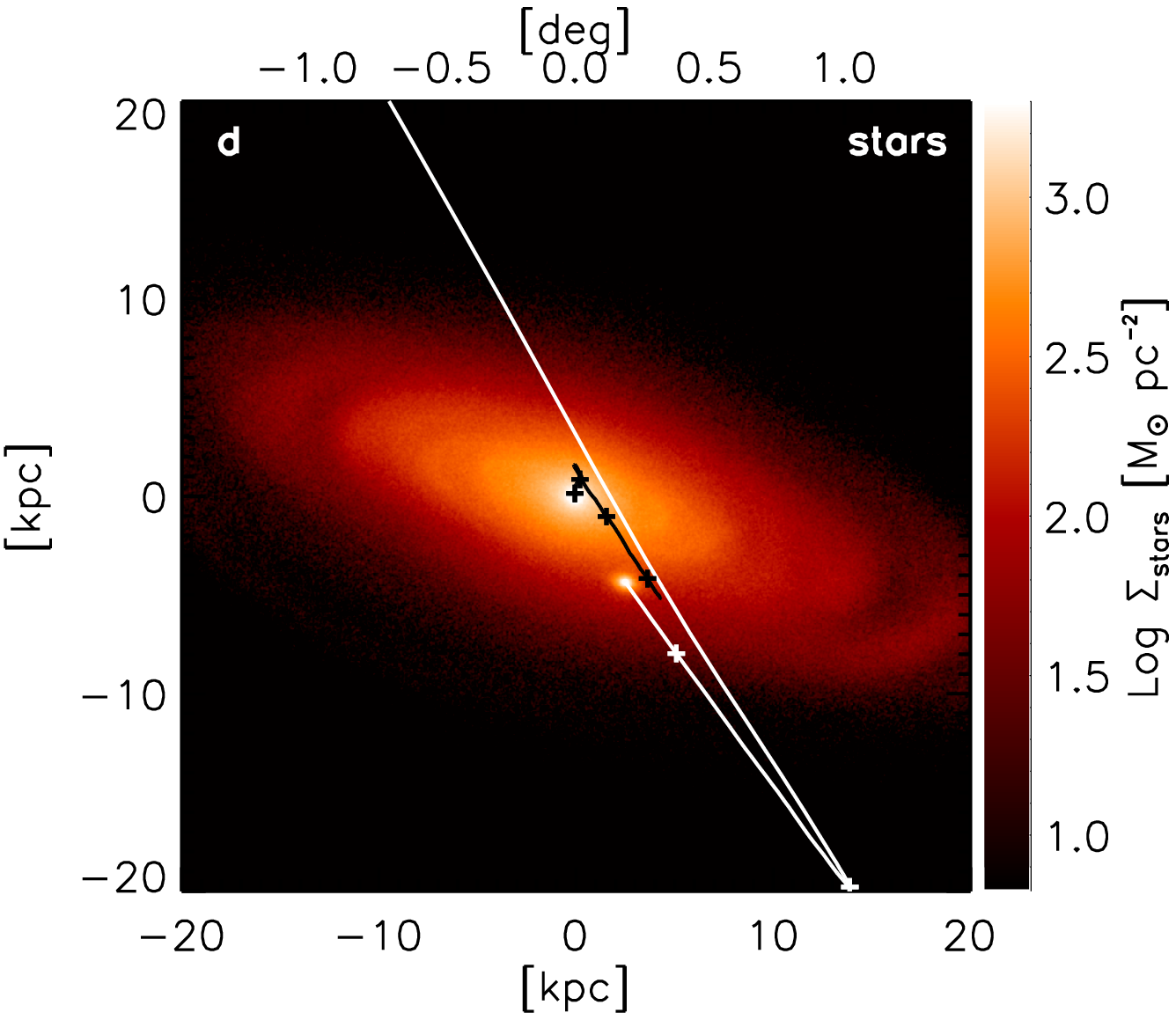}
\caption{ The simulated collision between Andromeda and M32 shown at the time of best match to current observations. {\bf Panels a and b:} gas morphology viewed face-on and in projection on the sky. {\bf Panel c:} face-on M32 dark matter density map (note the larger scale). {\bf Panel d:} stellar density map viewed in projection. In panel {\bf b}, a dashed ellipse marks the location of M31's 10~kpc pseudo-ring. The dim, incomplete outer ring tentatively identified in infrared images \citep{gordon06} is also reproduced. Black and white lines indicate the trajectories of Andromeda and M32, respectively, and include plus signs spaced every 500~Myr. Angular scales are calculated assuming a distance to M31 of 780~kpc. 
\label{fig:mosaic} }
\end{center}
\end{figure*}

\subsection{General features of the orbit}

 % Simulation generalities, M31 and M32 morphologies
Our final, high-resolution simulation of the best-match orbit has a baryonic (dark) particle mass resolution of $2\times 10^4$ M$_{\odot}$ ($5\times 10^5$ M$_{\odot}$). An overview of our fiducial simulation is presented in Figure~\ref{fig:mosaic}, and its parameters are summarized in Table~\ref{tab:params}. The observed gas and stellar morphologies of Andromeda are reproduced very well. The off-center impact generates expanding spiral arm structure, which in projection on the sky appears ring-like (``pseudo-rings"), bearing a strong resemblance to infrared maps \citep[e.g. \emph{Spitzer} MIPS in][]{gordon06,draine14}. Using a combination of infrared imaging, \citet{draine14} identify a deficiency of dust between $\sim$16 and 20~kpc on the southwest side of Andromeda's disk, a region also found to be deficient in HI gas \citep{nieten06, braun09}. Indeed, our contour density maps (Figure~\ref{fig:mosaic}b) are consistent with a below-average gas density in that sector. This localized deficiency could be related to the fact that M32 passes through the southwest side of the disk in our simulation.

M32's current observed position is reached 2~Gyr after the start of the simulation, shortly after apocenter. Because M32 and M31 are almost aligned along our line-of-sight and the simulated orbit is close to radial, the low observed M31-M32 relative radial velocity ($\sim100$~km~s$^{-1}$) constrains the orbit to be near turnaround. The current M31-M32 separation is therefore near maximum, with M32 lying $\sim85$~kpc in front of Andromeda in projection on the sky. Although the two galaxies lie close together in projection and are commonly assumed to be at essentially the same distance, our result is within the $2\sigma$ range of existing distance measurements \citep{freedman89,monachesi11,sarajedini12}. The prediction that M32 lies 10\% closer to us than M31 may be testable with improvements in the calibration of red giant branch distance indicators. 

Contrary to previous models, which place the collision less than $\sim$200~Myr ago \citep{block06,gordon06}, in our simulation the passage occurred 800~Myr in the past. This is consistent with the star-forming history of the pseudo-rings, which constrain the last major interaction between M31 and a companion to have occurred at least $\sim$500~Myr ago \citep{davidge12}. Our simulation also shows that a realistic passage with an inclined (non-polar) incidence angle and an intermediate impact parameter of 10~kpc leads to slightly offset pseudo-rings, as observed. A significant amount of the dark matter stripped from M32 has accreted onto M31, while the rest forms large-scale tidal features (Figure~1c).

\subsection{The collision's effect on the orbit of M31}

\begin{deluxetable}{lll}
\tablewidth{0pt}
\tablecaption{Velocity properties of the fiducial orbit.}
\tablehead{Velocity magnitude (km s$^{-1}$) & Andromeda (M31) & M32}
\startdata
Current full $v$ & 1.6 & 117  \\
Current $v_{\rm transverse}$ & 1.4 & 53 \\
\hline
Mean full $v$ over past 2 Gyr & 8.8 & 148\\
Mean $v_{\rm transverse}$ over past 2 Gyr & 5.0 & 100 \\
\hline
Max full $v$ over past 2 Gyr & 33 & 514 \\
Max $v_{\rm transverse}$ over past 2 Gyr & 16 & 361 \\
\enddata
\label{tab:velinfo}
\end{deluxetable}

Initially at rest in the simulation, M31's total displacement over the course of the interaction is $\sim15$~kpc. The projected shift of M31's disk on the sky is $\sim 0.5^\circ$ (see Figure~1), comparable to the angle subtended by the full Moon. Note that this could also be predicted analytically; in the point-particle limit, a total M31 displacement of twice the initial distance from the M31-M32 barycenter, or 18~kpc, is expected. This highlights the importance that minor dwarf satellites might have had in the dynamical history of the Local Group. 

Table \ref{tab:velinfo} presents kinematic information for the two simulated galaxies. M32 is predicted to have a current transverse velocity magnitude of $\sim$~50~km s$^{-1}$ with respect to the Milky Way. However, since the orbit is presently near turnaround, the current velocity signature on M31 is small; its present transverse speed is only $\sim 1.4$~km~s$^{-1}$. This is consistent with a radial orbit toward our Galaxy, in agreement with Andromeda-Milky Way head-on collision scenarios \citep{cox08,vdM12II}. Various indirect galactocentric transverse velocity estimates yield values of $V_{\rm tan} \simeq 17-100$~km~s$^{-1}$ \citep{vdM08,sohn12,vdM12II} with significant uncertainties. The discovery of five water masers in M31 \citep{darling11} is the first step toward a detailed proper motion study using Very Large Baseline Interferometry.

\subsection{Kinematic effects on Andromeda's disk}

% Spiral arm radial velocity signature
 \begin{figure*}[hbt]
\begin{center}
\includegraphics[scale=0.76,trim = 9mm 2mm 8mm 3mm, clip]{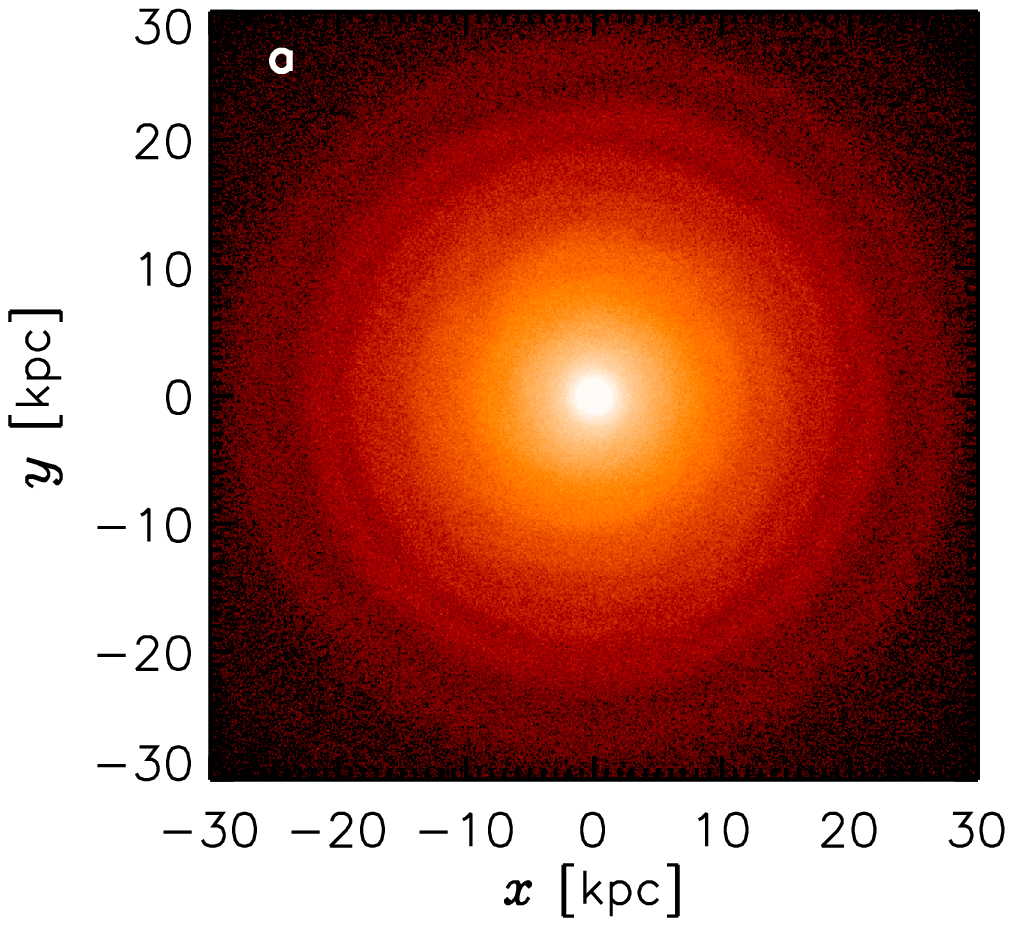}
\includegraphics[scale=0.65,trim = 9mm 0mm 0mm 3mm, clip]{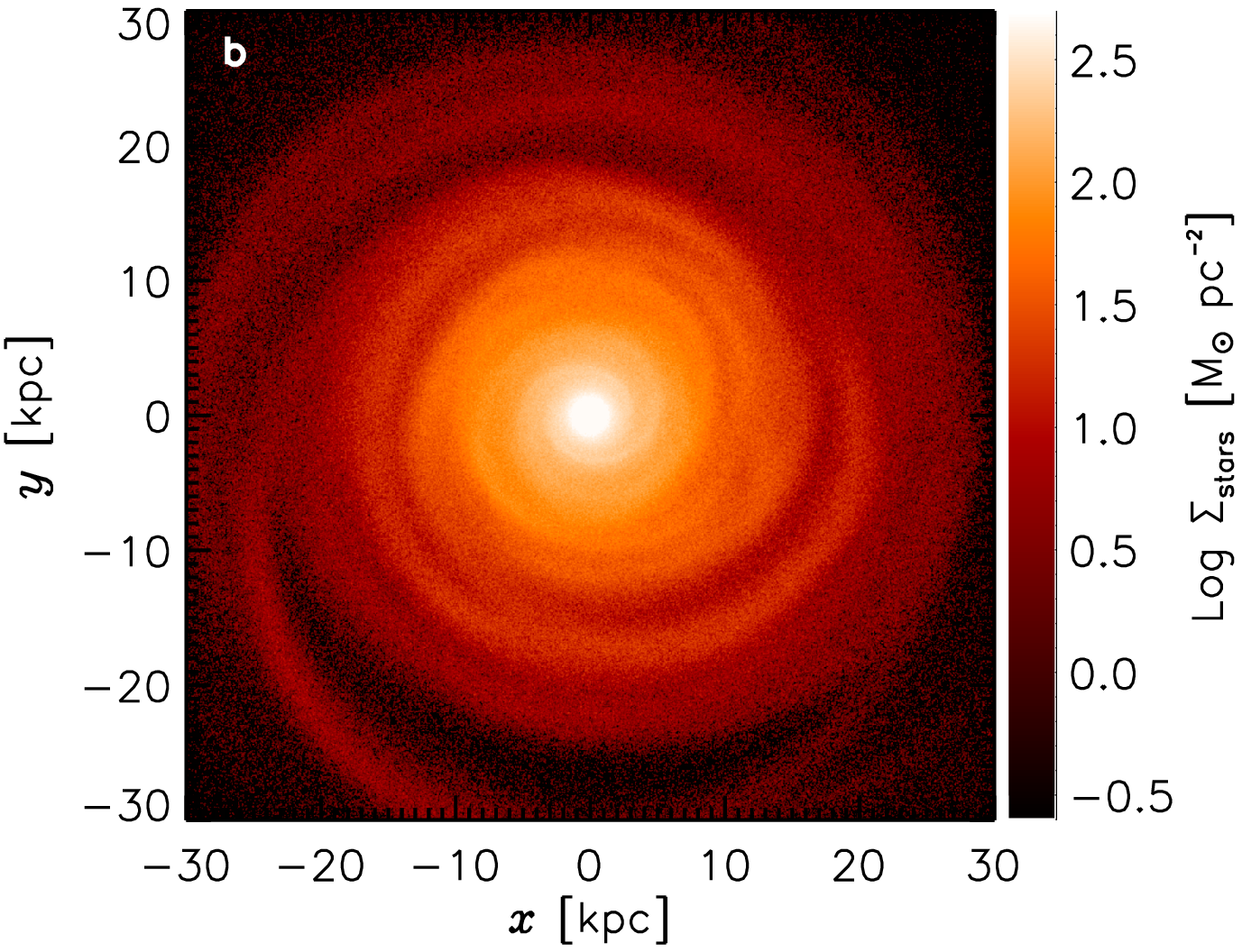}
\includegraphics[scale=0.865,trim = 0mm 14mm 19mm 13mm, clip]{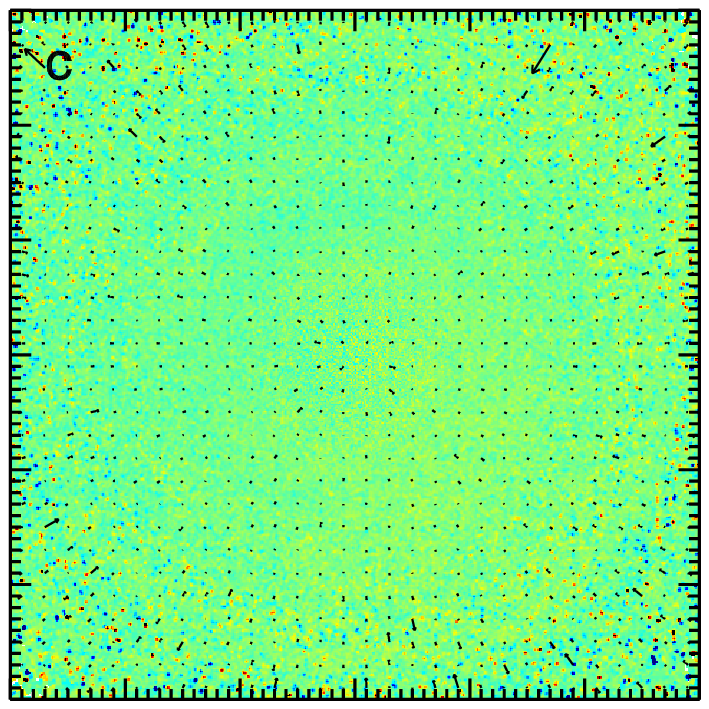}
\includegraphics[scale=0.865,trim = 0mm 14mm 8.5mm 7mm]{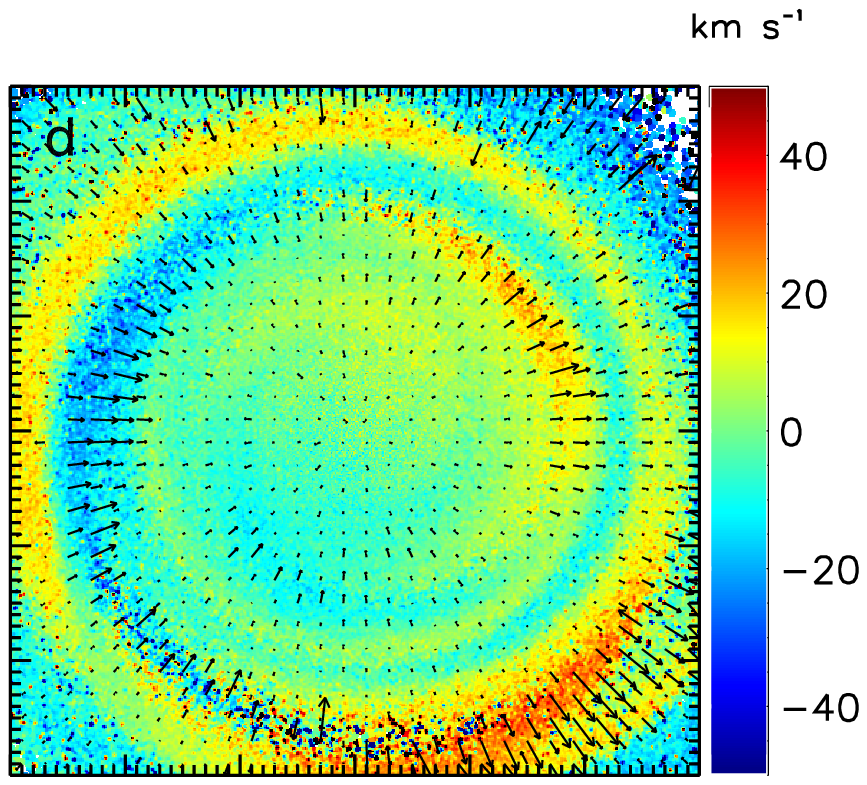}
\caption{Kinematic structure of Andromeda's stellar disk seen face-on.  {\bf Left panels:} disk evolved in isolation. {\bf Right panels:} disk after the interaction with M32 (prediction for the current time). The {\bf top row} shows the corresponding stellar density maps. The {\bf bottom row} presents the velocity structure of the stellar particles: velocity perpendicular to the disk plane (color scale) and in-plane component along the direction to M31's center (arrows). An arrow of length 1~kpc corresponds to a magnitude of 20~km~s$^{-1}$. Included here are all particles within three scale heights of the disk midplane. The data in the {\bf right panels} are rotated by -3$^\circ$ around the $x$-axis and 3$^\circ$ around the $y$-axis in order to compensate for the disk tilt induced by M32's passage. The pseudo-ring features produced by the collision are traced out by velocity excursions.
\label{fig:rvfield} }
\end{center}
\end{figure*}

% Radial velocity plots
\begin{figure*}[hbt]
\begin{center}
\includegraphics[scale=0.5,trim = 8mm 4mm 6mm 2mm, clip]{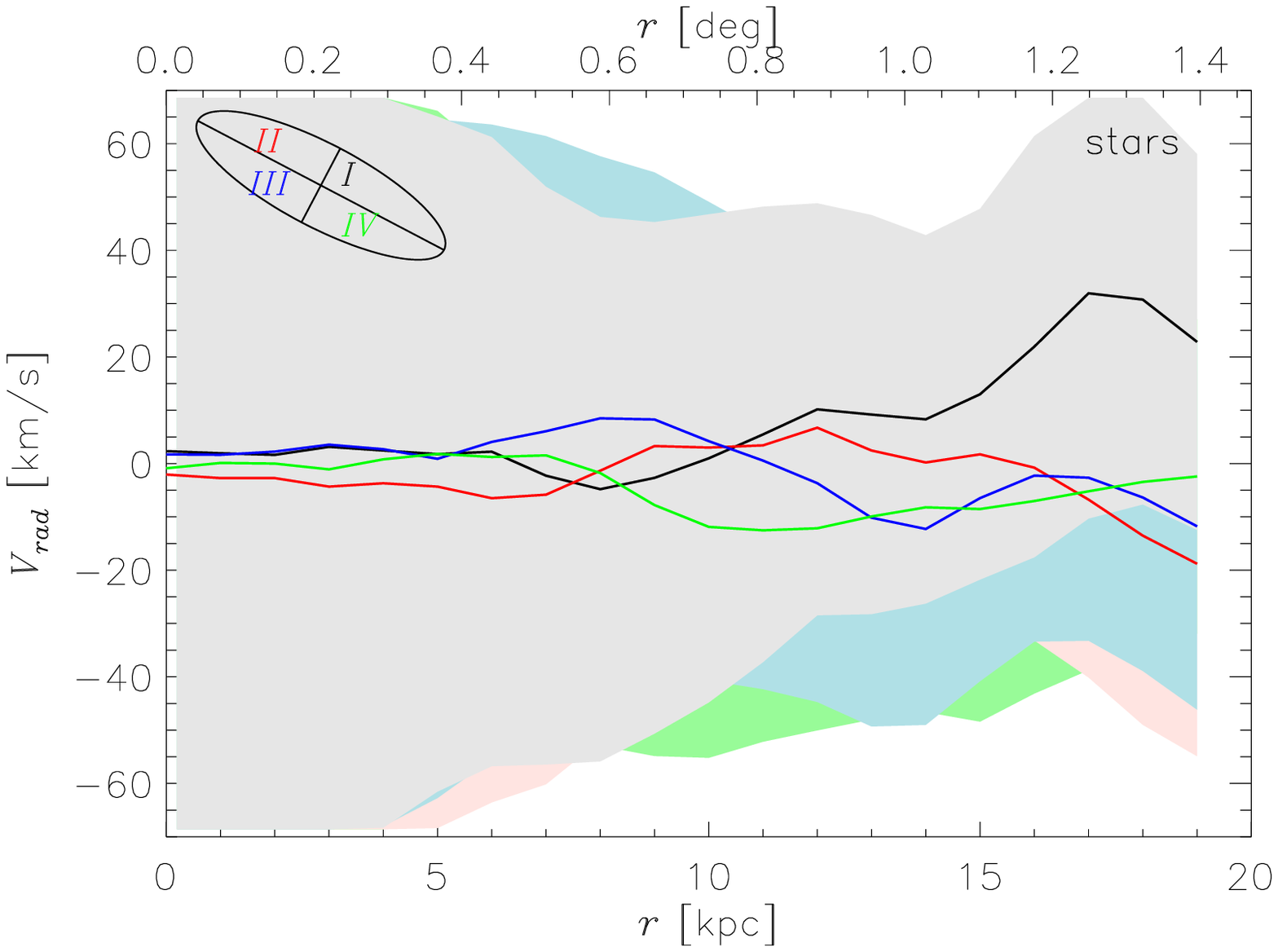}
\includegraphics[scale=0.5,trim = 8mm 4mm 6mm 2mm, clip]{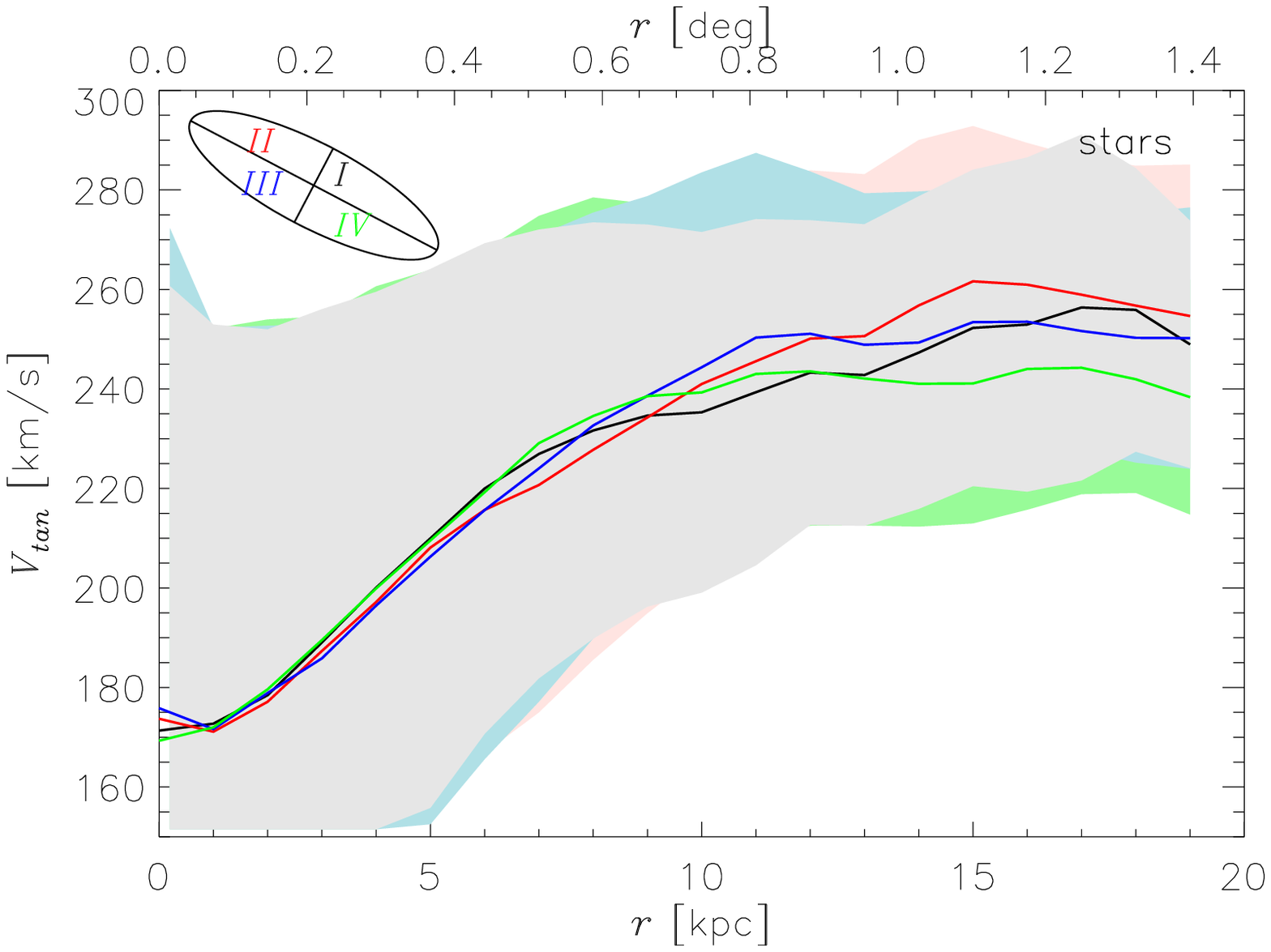}
\includegraphics[scale=0.5,trim = 8mm 4mm 6mm 2mm, clip]{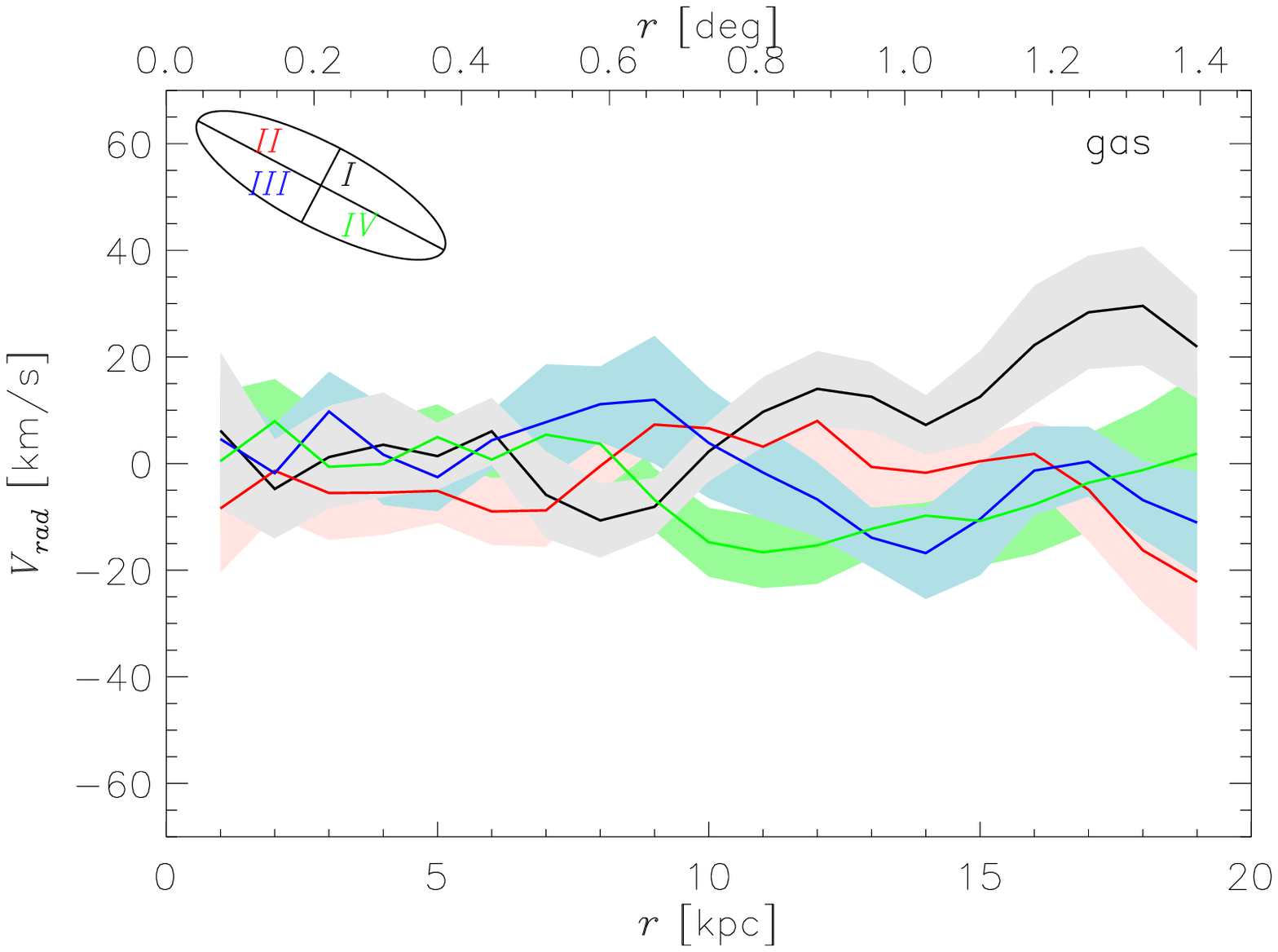}
\includegraphics[scale=0.5,trim = 8mm 4mm 6mm 2mm, clip]{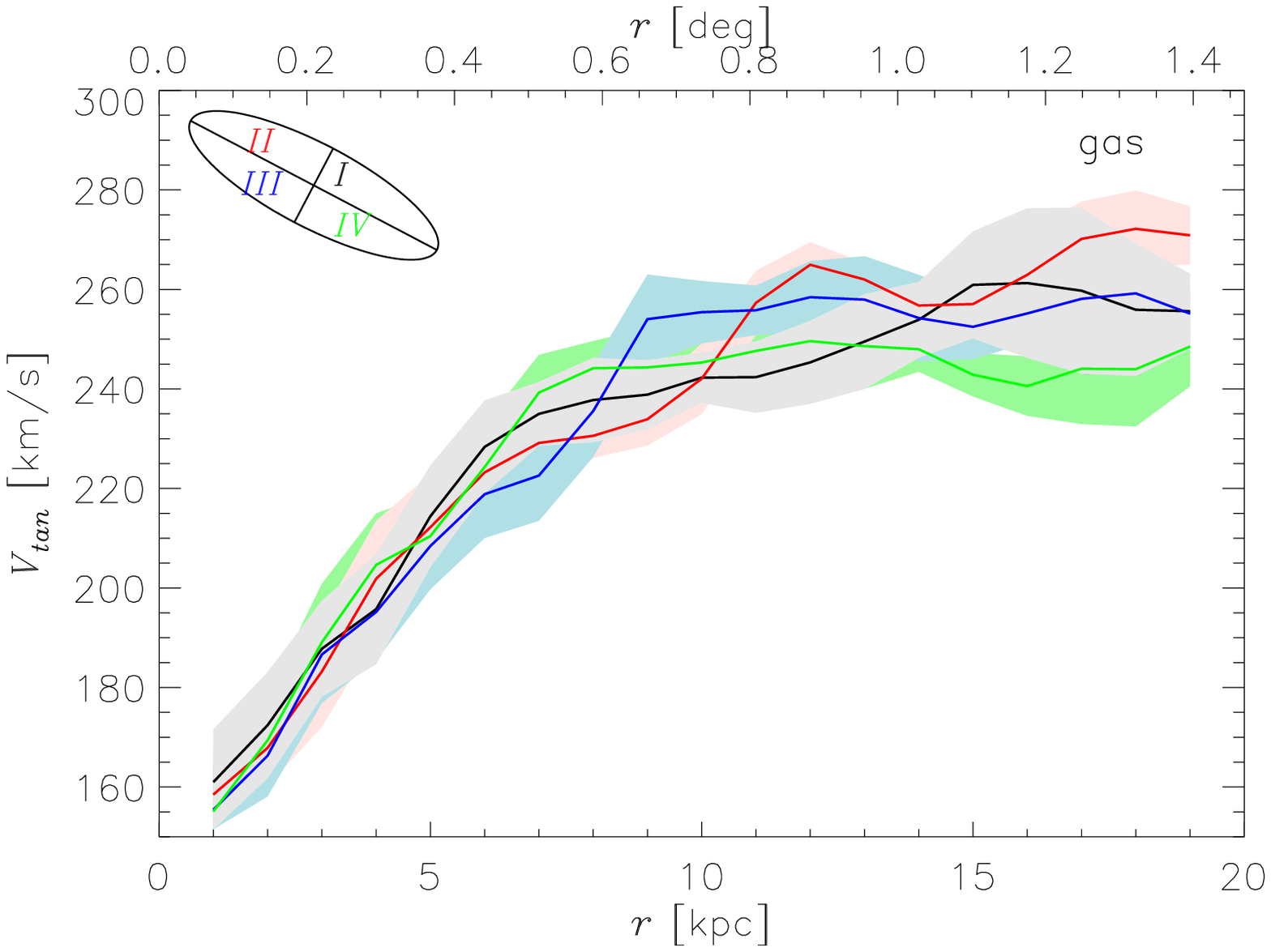}
\caption{Components of azimuthally-averaged star ({\bf top row}) and gas ({\bf bottom row}) particle velocities as a function of radius for different quadrants of Andromeda's disk in projection. The {\bf left panels} present the radial component of the velocities,  while the {\bf right panels} show the tangential component. As in Figure~\ref{fig:rvfield}, the data are rotated by -3$^\circ$ around the $x$-axis and 3$^\circ$ around the $y$-axis to compensate for the disk tilt induced by M32's passage. A color-quadrant location key is given in the top left corner of each panel. Colored bands correspond to the standard deviation of the velocity distribution in each quadrant. The increasing dispersion towards smaller radii in the stellar velocities is caused by the bulge component. Included here are all particles within three scale heights of the disk midplane. For the gas panels, the first radius bin is omitted as the density is too low near the center (a numerical artifact). Velocity excursions are related to the ring-like density enhancements seen in Figure~\ref{fig:mosaic}; the fact that they are not collocated in radius is expected because velocity extrema do not correspond to where particles collect in space. 
\label{fig:starsvelfield} }
\end{center}
\end{figure*}

Imperfections in the rotation curve modeling of Andromeda are expected to add a measurable uncertainty to proper motion observations \citep{darling11}. In Figure~\ref{fig:rvfield} we present a face-on map of Andromeda's stellar disk, showing the velocity perpendicular to the disk plane and the in-plane component along the direction to M31's center. A comparison with M31's disk simulated in isolation highlights the significant perturbations caused by M32's passage. The pseudo-rings map to $\sim$40~km~s$^{-1}$ deviations perpendicularly to the disk plane and in the radial direction. Observed deviations in maser transverse velocities are on the same order \citep{darling11}, suggesting that velocity perturbations induced by M32 may affect detailed modeling of M31's disk. 

In Figure \ref{fig:starsvelfield} we present the averaged radial and tangential components of the stellar and gas particle velocities relative to the disk center. For a simulation where M31 is evolved in isolation, deviations from a smooth velocity profile are only of order 5~km~s$^{-1}$, confirming that the perturbations in Figure~\ref{fig:starsvelfield} are caused by M32's passage. In the northwest quadrant (I) of Andromeda's disk, on average the particles located at a radius of 15-20~kpc have outwardly directed radial velocities peaking at $\sim$30~km~s$^{-1}$. This excursion could be related to the location of the simulated passage between sectors I and IV of the disk.

Finally, we note that Andromeda's disk grows in size on the sky as it approaches the Milky Way at $v_{\rm rad} \simeq 110$~km~s$^{-1}$. This apparent expansion can be approximated as $\dot{\theta}_{\rm app} \sim Rv_{\rm rad}/D^2$, where $R\sim20$~kpc is the physical radius of the disk and $D \simeq 780$~kpc the distance to Andromeda. Because the perturbations to the radial velocity profile due to M32's passage are of order $\Delta v_{\rm rad} \sim 30$~km~s$^{-1}$ (Figures \ref{fig:rvfield} and \ref{fig:starsvelfield}), the corresponding deviations in angular velocity, $\Delta \dot\theta_{\rm rad} \sim \Delta v_{\rm rad} / D$, dominate the apparent angular expansion by a factor of 10. Disentangling this effect from peculiar motions and apparent expansion represents a new complication for future maser proper motion studies. 

\subsection{Formation mechanism of M32}
\label{subsec:m32formation}

The formation mechanism of M32-like galaxies remains under debate. Usually found close to massive neighbors, cEs could be tidally stripped remnants, a scenario which does not explain the finding of an apparently isolated M32 twin \citep{huxor13}. From a theoretical standpoint, previous simulations \citep{bekki01,choi02} are limited by the treatment of the host galaxy as a static analytic potential over short timescales. 

In our simulations, only M32's first passage is disruptive enough to produce significant pseudo-rings in Andromeda. cE formation scenarios in which the satellite is stripped over many passages are therefore difficult to reconcile with ring formation via interaction with M32. Tidal stripping during this first passage decreases M32's total mass by more than a factor of ten, yielding a final mass of $\sim 2 \times 10^9$~M$_{\odot}$ consistent with current estimates \citep{nolthenius86,mateo98}. However, the passage primarily strips matter from the halo of M32 and reduces its half-light radius by only $\sim20-40$\%. The minimum tidal radius attained during the interaction is $\sim 1.8$~kpc, meaning that the inner baryonic component is not sufficiently stripped to yield a cE-like morphology. This finding is in agreement with surface photometry studies, which observe a lack of evidence of tidal stripping inside of $\sim 1$~kpc from M32's center \citep{choi02,howley13}. The fact that the bulge is left intact in our simulations is also consistent with the fact that M32 lies on the observed black hole mass~--~bulge stellar velocity dispersion relation for undisturbed bulges \citep{ferrarese00}. Our simulated M32 retains an exponential, gas-poor disk component qualitatively consistent with observations by \citet{graham02}. This suggests that M32, with its effective radius of $\sim 0.1$~kpc, may have evolved from an intrinsically compact progenitor and lends support to the idea that not all cEs are tidally-stripped remnants. An alternate scenario, where Andromeda's ring-like structure is caused by past interaction with a different satellite and M32's compact morphology comes from tidal stripping over multiple passages, cannot be excluded given current observed constraints. 

From interferometric observations of neutral and molecular hydrogen, an upper limit of only 8$\times10^4$~M$_\odot$ has been derived for the cool gas mass present within 1.3' of M32's center \citep{welch01}. M32's missing interstellar medium could be due to a combination of gas stripping, star formation and active galactic nuclei (AGN) feedback. However, 2-5~Gyr old stars contribute $\sim$40\% of the stellar mass in M32 and there is little evidence of star formation in the past 2~Gyr \citep{monachesi12}. Host to a $(2.4\pm1.0)\times10^6$~M$_\odot$ black hole \citep{vandenbosch10} shining at $\sim 2\times 10^{-8}$ the Eddington luminosity \citep{ho03}, M32's nucleus is currently quiescent, presumably due to its gas-poor environment. However, AGN feedback from past episodes of black hole growth may have played a role in removing gas from M32. Recent searches \citep[][and references therein]{reines13} have lead to an increased sample of dwarfs with known AGN. 

Here we discuss whether ram pressure stripping in our candidate orbit could lead to the observed lack of gas. From momentum conservation, $v_{\rm impact} \simeq 500$~km~s$^{-1}$ yields a velocity boost of 63~km~s$^{-1}$ for the M32 gas in the frame where it is initially at rest. Given a fiducial M32 circular velocity of 60~km~s$^{-1}$, $v_{\rm final} \gtrsim 150$~km~s$^{-1}$ is required to unbind the gas in the simulation. This suggests that the gas in our M32 model will not be completely unbound by the passage, in agreement with our fiducial simulation where only a small amount of stripping is observed. While the gas mass within one tidal radius decreases by a factor of 3, the mass inside of one disk scale length is only reduced by $\sim20$\%, leaving 50 times more gas than the observed upper limit of 8$\times10^4$~M$_\odot$ \citep{welch01}. For strong gas stripping, we find that an initial gas fraction of 0.01 is required, which we cannot adequately resolve given the baryonic mass resolution of our highest-resolution simulation. To summarize, we estimate that the gas in M32 can be efficiently removed by a single passage through M31's disk provided the progenitor is gas-poor (initial gas mass of less than $\sim 10^6$~M$_\odot$), but resolution limits prevent us from robustly testing this conclusion directly. The predicted collision with M31 could plausibly trigger gas removal via a combination of gas stripping, enhanced star formation and AGN activity.

%%%%%%%%%%%%%%%%%%%%%%%%%%%%%%
\section{Summary}
\label{sec:conclusions}

An off-center collision with dwarf companion M32 explains the apparent nested ring morphology of Andromeda's disk. Because the impact parameter of the collision is allowed to be a factor of ten larger than in previous works \citep[e.g.][]{block06}, the orbit is much more probable. Under this scenario, M32's passage occurred 800~Myr ago and produced measurable velocity perturbations in Andromeda's disk. The simulated orbit implies that M32 is $\sim100$~kpc closer to the Milky Way than previously thought. Our simulations are the first to model the evolution of the combined system over 2 billion years in a manner consistent with the observed positions and velocities of both galaxies. Only the first passage is disruptive enough to generate rings in M31, and the associated tidal stripping is insufficient to produce an M32-like morphology, supporting an intrinsically compact origin for cEs. 
Too many phase-space coordinates remain unknown to claim that this orbital solution is unique. However, if M32's passage is responsible for the pseudo-ring structure, our fiducial simulation must reproduce the general properties of the true M32 orbit. 
The lack of gas in M32 remains puzzling: for our fiducial orbit, ram pressure stripping becomes a viable removal mechanism only if M32's progenitor has a very low gas surface density, suggesting that M32 may have been gas-poor prior to the collision.

\acknowledgments
We thank Jeremy Darling for helpful comments on the manuscript. This work was supported in part by NSF grant AST-1312034. Support for LB was provided by NASA through the Einstein Fellowship Program, grant PF2-130093.

\end{document}